\documentclass[apjl]{emulateapj-rtx4}

\usepackage{times}
\usepackage{natbib}
\shorttitle{Excitation of Slow-Modes in Network Magnetic Elements}
\shortauthors{Yoshiaki Kato et al.}

\begin{document}

\title{Excitation of Slow-Modes in Network Magnetic Elements Through Magnetic Pumping}

\author{Yoshiaki Kato}
\affil{Institute of Space and Astronautical Science, Japan Aerospace
 Exploration Agency, 3-1-1 Yoshinodai, Chuo-ku, Sagamihara, Kanagawa
 252-5210, Japan}
\email{kato.yoshiaki@isas.jaxa.jp}
\author{Oskar Steiner}
\affil{Kiepenheuer-Institut f\"ur Sonnenphysik, Sch\"oneckstrasse 6,
      D-79104 Freiburg, Germany}
\author{Matthias Steffen}
\affil{Astrophysikalisches Institut Potsdam, An der Sternwarte 16, 
      D-14482, Potsdam, Germany}

\author{Yoshinori Suematsu}
\affil{Hinode Science Center, National Astronomical Observatory of Japan,
      2-21-1 Osawa, Mitaka, Tokyo 181-8588, Japan}

\begin{abstract}
From radiation magnetohydrodynamic simulations of the solar
atmosphere we find a new mechanism for the excitation of longitudinal 
slow modes within magnetic flux concentrations. We find that the convective 
downdrafts in the immediate surroundings of magnetic elements are responsible 
for the excitation of slow modes.
The coupling between the external downdraft and the plasma motion
internal to the flux concentration is mediated by the inertial forces
of the downdraft that act on the magnetic flux concentration. These
forces, in conjunction with the downward movement, 
pump the internal atmosphere in the downward direction, which
entails a fast downdraft in the photospheric and chromospheric layers
of the magnetic element. Subsequent to the transient pumping phase, the
atmosphere rebounds, causing a slow mode traveling along the magnetic 
flux concentration in the upward direction.  It develops into a shock wave 
in chromospheric heights, possibly capable of producing some kind of dynamic 
fibril.  We propose an observational detection of this process.
\end{abstract}


%%%%%%%%%%%%%%%%%%%%%%%%%%%%%%%%%%%%%%%%%%%%%%%%%%%%%%%%%%%%%%%%%%%%%%%%%%%%%%%%%%%

\section{Introduction}
\label{sect1}

Filtergrams or spectroheliograms in \ion{Ca}{2} H and K reveal two
main sources of \ion{Ca}{2} emission: plages, which spatially coincide
with the active regions, and the chromospheric network, which outlines
the boundaries of the supergranular velocity field
\citep{simon+leighton1964}.  Both are also the location of small
magnetic flux concentrations in the photosphere and dynamic fibrils 
in the chromosphere \citep{hansteen+al2006}.  The close
relationship between \ion{Ca}{2} emission and magnetic flux
\citep{skumanich+al1975,schrijver+al1989} suggests that the magnetic
field plays a key role for the chromospheric emission. 
The source of this emission has remained elusive, but likely
candidates are the dissipation of magnetohydrodynamic waves, or the
direct dissipation of electric currents. With regard to the former
process, numerous studies have been carried out based on
the approximation of slender flux tubes 
\citep[see, e.g.,][and references therein]{hasan+ulmschneider2004}.
They have greatly expanded
our understanding of the physics of tube modes, mode coupling,
dependency on the excitation spectrum and the tube geometry, shock
formation, etc. 

Some of these models consider the generation of tube waves by
turbulent motions in the convection zone, where an analytical
treatment of turbulence based on the Lighthill-Stein theory of sound
generation is used, e.g., in \citet{musielak+al1989,musielak+al1995,
 musielak+al2000} and \citet{musielak+ulmschneider2001}. Others
consider the generation of transverse waves through the impulse
transmitted by granules to magnetic flux tubes, as, e.g., in
\citet{choudhuri+al1993a,choudhuri+al1993b}, \citet{hasan+al2000}, and
\citet{cranmer+van_ballegooijen2005}.

Wave generation and propagation in thick tubes were considered by
\citet{ziegler+ulmschneider1997a,ziegler+ulmschneider1997b} who found
the development of a non-axisymmetric longitudinal body wave 
and energy leakage by acoustic radiation.  \citet{huang+al1999}
investigated surface and body waves of two-dimensional magnetic
slabs.  More recently \citet{rosenthal+al2002}, \citet{bogdan+al2003},
\citet{hasan+al2005}, \citet{hasan+vanBallegooijen2008},
\citet{khomenko+al2008}, and  \citet{vigeesh+al2009} carried out
numerical simulations of magneto-acoustic wave propagation for an
environment resembling the network atmosphere including small-scale
magnetic flux concentrations with internal structure. They found that
refraction of waves and mode conversion may play an important role in
the conversion of wave energy to heat in the network.

All the above mentioned models and simulations impose a given driving,
which is either monochromatic or impulsive, or derived from a
theoretical spectrum of turbulence, or from observations.  The focus
of the present investigation is the self-consistent excitation
of a thick magnetic flux concentration through the ambient convective
motion and specific consequences of this excitation.

%%%%%%%%%%%%%%%%%%%%%%%%%%%%%%%%%%%%%%%%%%%%%%%%%%%%%%%%%%%%%%%%%%%%%%%%%%%%%%%%%%%

\section{Numerical Method and Model Characteristics}
\label{sect2}

\begin{figure}
 \epsscale{1.1}
 \plotone{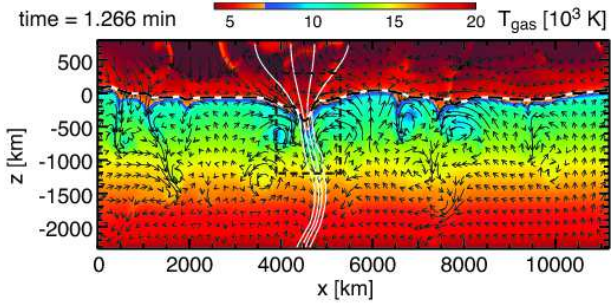}
 %\centerline{\epsfxsize=1.0\hsize \epsfbox{fig1.eps}}
 \caption{Snapshot of the simulation, showing the full computational
 domain. Color-scales show the gas temperature.  The white solid 
 curves are representative magnetic field lines, where the inner pair
 of field lines indicate the core region of the magnetic element and the
 outer pair of field lines the inner boundaries of the skin region. 
 The black and white dashed 
 curve shows the surface of optical depth unity, $\tau =1$. Arrows 
 indicate the plasma velocity. The black dashed box indicates the
 size of the close-up of Figure~\ref{fig_closeup}.}
\label{fig_synoptic}
\end{figure}

The simulations were carried out with the
CO\raisebox{0.5ex}{\footnotesize 5}BOLD-code 
\citep{freytag+al2002}. 
This code solves the coupled system of the equations of compressible
magnetohydrodynamics in an external gravity field and non-local,
frequency-dependent radiative transfer in one, two, or three spatial
dimensions. The main features of the code are described in
\citet{schaffenberger+al2005,schaffenberger+al2006}. The
two-dimensional computational domain extends over a height range of
3160\,km of which 780\,km reach above the mean surface of optical
depth unity. The horizontal extension is
11200\,km.  The grid-cell size in the horizontal direction is
28\,km, in the vertical direction it is 12~km in the upper part,
continuously increasing to 30~km through the convection-zone.  The
lateral boundary conditions are periodic in all variables, whereas the
lower boundary is open in the sense that the fluid can freely flow
in and out of the computational domain under the condition of
vanishing total mass flux. The specific entropy of the inflowing mass
is fixed to a value previously determined so as to yield the correct
solar radiative flux at the upper boundary. The upper
boundary is transmitting, viz., ${\rm d}v_{x,z}/{\rm d} z = 0$, 
while ${\rm d}\ln p_{\rm gas}/{\rm d} z = -1/H_{p}$ and 
${\rm d}\ln \rho/{\rm d} z = -1/H_{\rho}$ with $H_{p}$ and 
$H_{\rho}$ being the local gas pressure and density scale 
heights, respectively.
%while stress-free conditions are in effect for the horizontal
%velocities, viz., ${\rm d}v_{x}/{\rm d} z = 0$.

The simulation starts with a homogeneous, vertical, unipolar magnetic field 
of a flux density of 80~G superposed on a previously computed, relaxed
model of thermal convection. The magnetic field is constrained to have
vanishing horizontal components at the top and bottom boundary but
lines of force can freely move in the horizontal direction.  At the
beginning, the magnetic field quickly concentrates in the
intergranular downdrafts of the convective flow.   Subsequently,
individual flux concentrations merge and after 
$\approx 100$~minutes, the magnetic field concentrates in a single magnetic
`flux sheet' with a strength of 2000 to 2600~G at the optical
depth unity within the flux concentration.  This instant corresponds 
to the time $t = 0$ in our investigation. Subsequently, the magnetic field 
remains concentrated in this single magnetic element for the remaining 
86~minutes of simulation time.  

This state, however, is not a stationary one---as a consequence of the
interaction with the surrounding convective motion, the flux
concentration moves laterally, gets distorted, and exhibits internal
plasma flow. This interaction excites magneto-acoustic waves within
the flux sheet, in particular longitudinal slow modes. In the
following we point to a new mechanism of wave generation, which
was not considered before.

%%%%%%%%%%%%%%%%%%%%%%%%%%%%%%%%%%%%%%%%%%%%%%%%%%%%%%%%%%%%%%%%%%%%%%%%%%%%%%%%%%%

\section{Ambient downflows as a source of slow modes in magnetic elements}
\label{sect3}

Figure~\ref{fig_synoptic} shows the complete computational domain with
the magnetic flux concentration (white magnetic field lines) in the
middle, the temperature field (color scale), the velocity field
(arrows), and the continuum optical depth unity (dashed contour) 
for an arbitrary instant in time. The snapshot
shows warm granular upwellings framed by narrower cool intergranular
downdrafts in the convection zone and a few shock waves in the top
part of the atmosphere. It also shows two narrow downflow channels in
the close vicinity outside of the magnetic flux concentration around
and below the surface of optical depth unity. These `downflow jets'
are a consequence of a baroclinic flow impinging on the magnetic flux
concentration from the lateral directions, driven by the radiative
cooling at the `hot walls' of the flux concentration 
\citep{deinzer+al1984b, steiner+al1998}. 
Observational evidences for downflows at the edges of small scale 
magnetic flux concentrations were reported by \citet{rimmele2004}, 
\citet{ishikawa+al2007}, \citet{langangen+al2007}, and \citet{rezaei+al2007}.

\begin{figure}
\epsscale{1.1}
\plotone{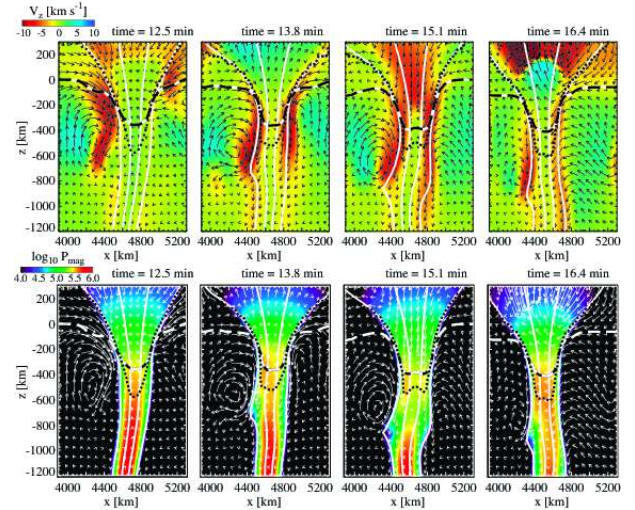}
%\centerline{\epsfxsize=0.95\hsize \epsfbox{fig2.eps}}
\caption{Time series of close-up views of the magnetic element, 
 demonstrating the pumping by external downflows.  Colors in the upper 
 panels show the vertical velocity, colors in the lower panels the magnetic 
 pressure. The white solid curves are representative field lines, where
 the inner pair of field lines indicate the core region and the outer 
 pair the inner boundaries of the skin region. The dashed curves
 correspond to the surfaces of optical depth unity, $\tau =1$, the
 dotted curves to surfaces of equal magnetic and gas pressure, $\beta =1$, 
 above which $\beta < 1$.  Negative velocities indicate downflows.
 %Note that the initially upflowing plasma  within
 %the magnetic element is pumped in the downward direction by action
 %of the fast ambient downdrafts.
}
\label{fig_closeup}
\end{figure}

In the present simulation, we observe that these downflows are far from
stationary. They tend to be present most of the time but get
transiently enhanced, weakened, or interrupted.  Sometimes, the
lateral inflow carries a preexisting regular intergranular downflow
with it. It then merges with the preexisting downflow channel of the flux
concentration, which results in a particularly strong downflow. 
Such advection of intergranular lanes to the downdrafts close
to magnetic flux concentrations are reported to exist from 
observations of faculae by \citet{depontieu+al2006}. In the following,
we identify such transients as a source of magneto-acoustic 
waves within the magnetic flux concentration, in particular
of longitudinal slow modes. 

\subsection{The magnetic pumping process}
\label{sect3:1}
Figure~\ref{fig_closeup} shows a close-up of a transient downflow and the
ensuing excitation of a slow wave over a time period of 234~s. The colors
represent the vertical velocity in the top row and the magnetic pressure
in the bottom row. The arrows indicate the flow speed, the vertically
running white curves selected magnetic field lines of the magnetic flux
concentration. 

At the beginning of the time sequence, there exists on the 
left side of the flux concentration a vortical flow, which is directed 
towards the magnetic flux concentration near optical depth 
unity, turning into a fast downflow immediately adjacent to the flux 
concentration.  A similar inflow and downflow exists on the 
right hand side of the flux concentration. In the course of time, 
these flow cells drift in the downward direction. The flows exert 
inertial forces on the magnetic flux sheet, which can be seen from
the sequence of the magnetic pressure, $p_{\rm mag}$. 

Adjacent to where the flow impinges on the flux sheet, the magnetic
pressure gets enhanced (e.g., at $z\approx -300$~km at time 13.8 minutes
and at $z\approx -500$~km at time 15.1 minutes),
while further downstream, where the flow speed is highest, and where
the flow starts to detach from the flux-sheet boundary, the flux sheet
expands and the magnetic pressure decreases. This situation leads to
an inversion in $p_{\rm mag}$ as a function of height, which moves in 
the downward direction together with the surrounding, gradually downward 
moving vortical flow cells. It presses material within the flux
sheet in the downward direction.
More illustratively expressed, the surrounding flow is (unidirectionally) 
kneading the flux sheet in the downward direction. This process entails 
a downflow within the magnetic flux concentration---while there exists 
a general upflow at time $t=12.5$~minutes, it turns into a downflow, 
first in the height range
where the kneading is strongest (around $\tau = 1$), later
gradually comprising the upper atmospheric layers up to the top of the
photosphere at time $t=15.1$~minutes.  The mechanism by which
convective flows in the surrounding of a magnetic flux tube may
induce a unidirectional flow within the tube was earlier recognized
and described by \citet{parker1974}, who called this process
`turbulent pumping'.\footnote{Note that nowadays, this term commonly
 refers to the transport of magnetic fields by turbulence in
 stratified media, which is a different process. We therefore
 call this process `magnetic pumping'.}
The principle is used in technical devices such as the peristaltic 
pump. \citet{parker1974} thought this process to be responsible
for the intensification of magnetic flux concentrations.
Here we find it to be responsible for the excitation of a
magneto-acoustic slow mode.
%at work in a numerical simulation of a magnetic flux sheet subject 
%to interaction with a convectively unstable environment. 

\subsection{The generation of the slow mode}
\label{sect3:2}
At first, the kneading process transports all material in the 
downward direction, but this process is not efficient enough to
prevent that some material can escape this trend and start to 
flow upwards in the center of the flux sheet, which is less directly 
subject to the surrounding flow.  This upflow starts around $\tau = 1$
at time $t=15.1$~minutes, while in the photosphere, plasma still flows
at high speed in the downward direction.  The high speed downflow
collides with the upflowing material of higher density, which leads to
an upwardly propagating wave.  Since this takes place in the low
 $\beta$ regime, and since the wave is guided along the magnetic
 field, it is a slow, essentially
acoustic wave. Further up it steepens into a shock front, visible 
at time $t=16.4$~minutes and $z\approx 200$~km.

\begin{figure}
\epsscale{0.9}
\plotone{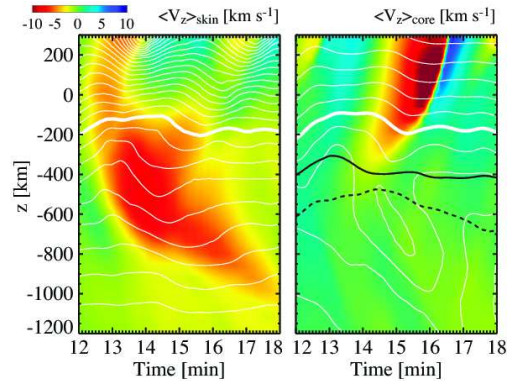}
%\centerline{\epsfxsize=0.8\hsize \epsfbox{fig3.eps}}
\caption{Space-time diagram of the close-up of Figure~\ref{fig_closeup}
 between times $t=12$~minutes and $t=18$~minutes. Left: Skin region of the
 magnetic element. Colors refer to the vertical velocity, white 
 contours to the logarithm of the gas pressure. 
 Right: Core region of the magnetic element. Colors refer to the vertical 
 velocity, white contours to the logarithm of the magnetic pressure. 
 Negative velocities indicate downflows.
 Distances between white contours are equal in both panels.  The
   gas pressure of the heavy white contour in the left panel is equal
   to the magnetic pressure of the heavy white contour in the right
   panel. Note that $\beta$ is not constant and very different for
   these two contours.
 The solid black
 curve indicates optical depth unity, the black dashed curve indicates
 the surfaces of equal gas pressure and magnetic pressure, 
 $\langle\beta\rangle_{\rm core}=1$. 
 The strong downdraft in the skin region (red region) entails 
 the downdraft in the magnetic element with subsequent shock formation 
 starting at about $(t,z) = (16\,{\rm min.},-50\,{\rm km})$.
 }
\label{fig_spacetime}
\end{figure}

Figure~\ref{fig_spacetime} shows two space-time diagrams of the event
depicted in Figure~\ref{fig_closeup}. The left panel shows the vertical
velocity (colors) and the gas pressure (contours) horizontally averaged 
over a 50~km thick skin region immediately outside of the magnetic flux 
sheet. The right panel shows the vertical velocity (colors) and the magnetic 
pressure (contours), horizontally averaged over a core region of the 
magnetic flux sheet, which encompasses one third of the total magnetic 
flux. The height $z=0$ corresponds to the mean optical depth unity outside
of the flux sheet.

A downdraft in the flux-sheet surroundings (left panel) starts at a 
height of $z\approx 0$~km and propagates downwards while strengthening: it also 
entails some replenishment from photospheric layers. The downdraft front
is accompanied by an enhancement in gas pressure, which later
turns into a gas pressure deficit as a consequence of Bernoulli's 
principle. Within the flux sheet (right panel), a downdraft ensues, first at
time $t\approx 13.3$~minutes closely below the 
$\langle\tau\rangle_{\rm core}=1$ level
at height $z\approx -400$~km. This downdraft cannot be 
replenished from the lateral direction (unlike in the skin region) 
but is forced to flow along the magnetic 
field from higher layers. Consequently, a fast downflow ensues from 
$z\approx -300$~km upwards. This downflow soon rebounds and
turns into a upflow, developping into a shock front, visible from 
$(t,z)=(16\,\mathrm{min.},-50\,\mathrm{km})$ to
$(t,z)=(16.6\,\mathrm{min.},+300\,\mathrm{km})$. 
At $z = 300$~km, the shock front moves with a speed of
approximately 15 km\,s$^{-1}$.
%, which is a plausible initial 
%velocity of a dynamic fibril \citep{depontieu+al2007}. 
The Alfv\'en speed at this height is $\approx 200$~km\,s$^{-1}$.

\begin{figure}
\epsscale{1.1}
\plotone{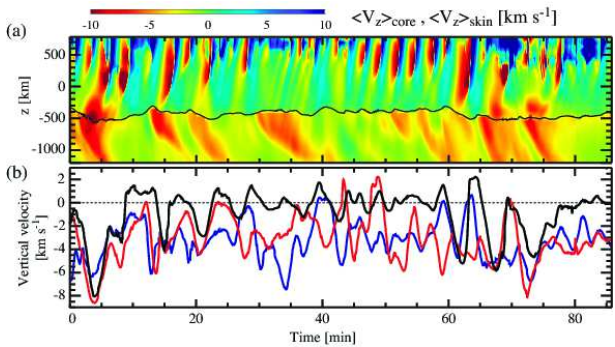}
%\centerline{\epsfxsize=0.8\hsize \epsfbox{fig4.eps}}
\caption{Top panel: Space-time diagram over the full time span of the
 simulation.  The black solid curve corresponds to the
 $\langle\tau\rangle_{\rm core}=1$ surface in the core of the flux
 sheet.  Colors refer to the vertical velocity in the core region of
 the magnetic element above $\langle\tau\rangle_{\rm core}=1$ and in
 the skin region of the magnetic element below
 ${\langle\tau\rangle}_{\rm core}=1$. Bottom panel: Vertical velocity
 spatially averaged between $-500\,{\rm km}\leq z\leq 0\,{\rm km}$ 
 as a function of time. The red curve refers  to the skin region on the
 right hand side of the magnetic element, the blue curve to the skin
 region on the left hand side. The black curve refers to the core
 region of the magnetic element.  Negative velocities indicate 
 downflows.}
\label{fig_spacetime_full}
\end{figure}

Figure~\ref{fig_spacetime_full}(a) shows the space-time diagram 
of the vertical velocity for the full time period of 86~minutes. The colors
above the height level of continuum optical depth unity,
${\langle\tau\rangle}_{\rm core} =1$, refer to the vertical 
velocity in the core region of the
flux concentration. Below $\langle\tau\rangle_{\rm core}=1$, they refer to 
$v_z$ in the skin region. 
$z=0$ corresponds to the optical depth unity outside of the flux
sheet, ${\langle\tau\rangle}_{\rm ext}=1$.  From this diagram it becomes
evident that some of the most vigorous downflows in the photospheric
layers of the flux sheet, 
e.g., at times $t=5.5$, 8.8, 15.1, 21.4, 62.8, 68.8, and
77.5~minutes, are causally related to precursory downflows in the skin
region of the flux sheet. 
The downflows in the skin region start at around optical depth unity. 
This is also visible in Figure~\ref{fig_spacetime_full}(b), which
shows the vertical velocity in the skin region to the left (blue) and to
the right (red) of the flux sheet at a height of around $z=0$. The solid black
curve gives the vertical velocity within the flux sheet at the same height.
Strong downflows and subsequent shock formation in the photospheric
layer of the magnetic flux concentration are often preceded by convective 
downflows in the close surroundings outside of the flux concentration.

From Figure~\ref{fig_spacetime_full}(b) one can see that most of the time there 
exists a downdraft in the close surroundings of the magnetic flux concentration.
However, strong downflow events occur sporadically only because they are triggered
by the unsteady convective motion. Such sporadic events are 
capable of exciting a non-linear, longitudinal wave within the magnetic flux 
concentration as a consequence of the pumping effect. 
In the wake of this impulsive excitation, the atmosphere within 
the flux sheet oscillates at the acoustic cut-off period 
\citep{rae+roberts1982,hasan+kalkofen1999}. This is the reason for the 
predominantly periodic pattern in the upper part of 
Figure~\ref{fig_spacetime_full}(a), despite of the sporadic, intermittent
excitation by the ambient downflows.

%%%%%%%%%%%%%%%%%%%%%%%%%%%%%%%%%%%%%%%%%%%%%%%%%%%%%%%%%%%%%%%%%%%%%%%%%%%%%%%%%%%

\section{Conclusions and Discussions}
\label{sect4}

We have carried out a two-dimensional MHD simulation of a network magnetic
element embedded in a non-stationary environment consisting of a 2.38~Mm thick
convection zone layer, and photospheric and chromospheric layers up to 0.78~Mm
above the optical depth $\tau = 1$.  In these simulations we found  a new
mechanism for the generation of longitudinal slow modes, which was not
considered before.  Sporadic strong downflows in the close
surroundings of the magnetic element,
%earlier seen and described in simulations
%\citep{steiner+al1998} and observations 
%\citep{rimmele2004,ishikawa+al2007,langangen+al2007},
`pump' the plasma within the magnetic flux element in the downward
direction by the action of inertial forces on the magnetic field---a
process that was first described by \citet{parker1974} who referred to
it as `turbulent pumping'. As soon as the transient ambient downflow 
weakens, the pumping comes to a halt and the downflowing plasma from
the photospheric and chromospheric layers of the magnetic element
rebounds, which leads to a slow, upwardly propagating magneto-acoustic
wave. It develops into a shock wave in the chromospheric layers of the 
magnetic element. In the wake of a `pumping event', the flux-tube atmosphere 
tends to oscillate at the acoustic cutoff-period. We identify these pumping 
events as a crucial mechanism for the excitation of longitudinal slow
modes in flux-tube atmospheres, the heating in network and plage areas, and
possibly the development of some kind of dynamic fibrils.

A major shortcoming of the present investigation is the restriction to
two spatial dimensions, which might favor the simultaneous presence of 
downflows on both sides of the flux sheet. At this point
we do not yet know to what extent the pumping mechanism carries over
to a three-dimensional environment. Also, we cannot exclude that
viscous forces (stemming from the implemented subgrid turbulence
model) may add to some extent to the pumping in our simulation.
However, the viscosity effect is judged minor as there is a rather 
sharp discontinuity in velocity across the $\beta=1$ contour, as can 
be seen from Figure~\ref{fig_closeup}.

For a future observational detection of the magnetic pumping 
mechanism, we propose 
to measure Doppler velocities within and in the close surroundings 
of magnetic elements at highest possible spatial resolution, and
for different spectral lines, similar to \citet{langangen+al2007}
  or \citet{bellot-rubio+al2001}. 
Different from these observations however, the measurements 
should preferably result in bi-dimensional Doppler maps, which would 
enable to 
keep track of the magnetic element. They should be supplemented 
with G-band and \ion{Ca}{2} H/K filtergrams for the identification 
of magnetic elements and the detection of the chromospheric response, 
respectively. In a time series of such measurements, we would expect 
a transient downdraft in the immediate surroundings of a magnetic
element to entail a downflow in the photospheric layers within the
magnetic element, followed by a fast upflow and a brightening in
\ion{Ca}{2} H/K as a consequence of the shock wave.

\begin{acknowledgements}
O.S.~gratefully acknowledges financial support and gracious
hospitality during his visiting professorship at the National
Astronomical Observatory of Japan (NAOJ), when part of the work
reported herein was carried out.  This work was supported in part by
the JSPS fund \#R53 (``Institutional Program for Young Researcher
Overseas Visits'', FY2009-2011) allocated to NAOJ, part of which is
managed by Hinode Science Center, NAOJ.  The numerical simulations were
carried out on the NEC SX-9 computer at JAXA Supercomputer Systems
(JSS). We are grateful to R.~Hammer for insightful discussions and to the referee for very helpful comments.\\
\end{acknowledgements}

%%%%%%%%%%%%%%%%%%%%%%%%%%%%%%%%%%%%%%%%%%%%%%%%%%%%%%%%%%%%%%%%%%%%%%%%%%%%%%%%%%%

%\bibliography{ms}

\begin{thebibliography}{36}
\expandafter\ifx\csname natexlab\endcsname\relax\def\natexlab#1{#1}\fi

\bibitem[{{Bellot Rubio} {et~al.}(2001){Bellot Rubio}, {Rodr{\'{\i}}guez
  Hidalgo}, {Collados}, {Khomenko}, \& {Ruiz Cobo}}]{bellot-rubio+al2001}
{Bellot Rubio}, L.~R., {Rodr{\'{\i}}guez Hidalgo}, I., {Collados}, M.,
  {Khomenko}, E., \& {Ruiz Cobo}, B. 2001, \apj, 560, 1010

\bibitem[{{Bogdan} {et~al.}(2003){Bogdan}, {Carlsson}, {Hansteen}, {McMurry},
  {Rosenthal}, {Johnson}, {Petty-Powell}, {Zita}, {Stein}, {McIntosh}, \&
  {Nordlund}}]{bogdan+al2003}
{Bogdan}, T.~J., {et~al.} 2003, \apj, 599, 626

\bibitem[{{Choudhuri} {et~al.}(1993{\natexlab{a}}){Choudhuri}, {Auffret}, \&
  {Priest}}]{choudhuri+al1993a}
{Choudhuri}, A.~R., {Auffret}, H., \& {Priest}, E.~R. 1993{\natexlab{a}},
  \solphys, 143, 49

\bibitem[{{Choudhuri} {et~al.}(1993{\natexlab{b}}){Choudhuri}, {Dikpati}, \&
  {Banerjee}}]{choudhuri+al1993b}
{Choudhuri}, A.~R., {Dikpati}, M., \& {Banerjee}, D. 1993{\natexlab{b}}, \apj,
  413, 811

\bibitem[{{Cranmer} \& {van Ballegooijen}(2005)}]{cranmer+van_ballegooijen2005}
{Cranmer}, S.~R., \& {van Ballegooijen}, A.~A. 2005, \apjs, 156, 265

\bibitem[{{De Pontieu} {et~al.}(2006){De Pontieu}, {Carlsson}, {Stein}, {Rouppe
  van der Voort}, {L{\"o}fdahl}, {van Noort}, {Nordlund}, \&
  {Scharmer}}]{depontieu+al2006}
{De Pontieu}, B., {Carlsson}, M., {Stein}, R., {Rouppe van der Voort}, L.,
  {L{\"o}fdahl}, M., {van Noort}, M., {Nordlund}, {\AA}., \& {Scharmer}, G.
  2006, \apj, 646, 1405

\bibitem[{{Deinzer} {et~al.}(1984){Deinzer}, {Hensler}, {Sch\"ussler}, \&
  {Weisshaar}}]{deinzer+al1984b}
{Deinzer}, W., {Hensler}, G., {Sch\"ussler}, M., \& {Weisshaar}, E. 1984, \aap,
  139, 435

\bibitem[{{Freytag} {et~al.}(2002){Freytag}, {Steffen}, \&
  {Dorch}}]{freytag+al2002}
{Freytag}, B., {Steffen}, M., \& {Dorch}, B. 2002, Astron.~Nachr., 323, 213

\bibitem[{{Hansteen} {et~al.}(2006){Hansteen}, {De Pontieu}, {Rouppe van der
  Voort}, {van Noort}, \& {Carlsson}}]{hansteen+al2006}
{Hansteen}, V.~H., {De Pontieu}, B., {Rouppe van der Voort}, L., {van Noort},
  M., \& {Carlsson}, M. 2006, \apjl, 647, L73

\bibitem[{{Hasan} \& {Kalkofen}(1999)}]{hasan+kalkofen1999}
{Hasan}, S.~S., \& {Kalkofen}, W. 1999, \apj, 519, 899

\bibitem[{{Hasan} {et~al.}(2000){Hasan}, {Kalkofen}, \& {van
  Ballegooijen}}]{hasan+al2000}
{Hasan}, S.~S., {Kalkofen}, W., \& {van Ballegooijen}, A.~A. 2000, \apjl, 535,
  L67

\bibitem[{{Hasan} \& {Ulmschneider}(2004)}]{hasan+ulmschneider2004}
{Hasan}, S.~S., \& {Ulmschneider}, P. 2004, \aap, 422, 1085

\bibitem[{{Hasan} \& {van Ballegooijen}(2008)}]{hasan+vanBallegooijen2008}
{Hasan}, S.~S., \& {van Ballegooijen}, A.~A. 2008, \apj, 680, 1542

\bibitem[{{Hasan} {et~al.}(2005){Hasan}, {van Ballegooijen}, {Kalkofen}, \&
  {Steiner}}]{hasan+al2005}
{Hasan}, S.~S., {van Ballegooijen}, A.~A., {Kalkofen}, W., \& {Steiner}, O.
  2005, \apj, 631, 1270

\bibitem[{{Huang} {et~al.}(1999){Huang}, {Musielak}, \&
  {Ulmschneider}}]{huang+al1999}
{Huang}, P., {Musielak}, Z.~E., \& {Ulmschneider}, P. 1999, \aap, 342, 300

\bibitem[{{Ishikawa} {et~al.}(2007){Ishikawa}, {Tsuneta}, {Kitakoshi},
  {Katsukawa}, {Bonet}, {Vargas Dom{\'{\i}}nguez}, {Rouppe van der Voort},
  {Sakamoto}, \& {Ebisuzaki}}]{ishikawa+al2007}
{Ishikawa}, R., {et~al.} 2007, \aap, 472, 911

\bibitem[{{Khomenko} {et~al.}(2008){Khomenko}, {Collados}, \&
  {Felipe}}]{khomenko+al2008}
{Khomenko}, E., {Collados}, M., \& {Felipe}, T. 2008, \solphys, 251, 589

\bibitem[{{Langangen} {et~al.}(2007){Langangen}, {Carlsson}, {Rouppe van der
  Voort}, \& {Stein}}]{langangen+al2007}
{Langangen}, {\O}., {Carlsson}, M., {Rouppe van der Voort}, L., \& {Stein},
  R.~F. 2007, \apj, 655, 615

\bibitem[{{Musielak} {et~al.}(1995){Musielak}, {Rosner}, {Gail}, \&
  {Ulmschneider}}]{musielak+al1995}
{Musielak}, Z.~E., {Rosner}, R., {Gail}, H.~P., \& {Ulmschneider}, P. 1995,
  \apj, 448, 865

\bibitem[{{Musielak} {et~al.}(1989){Musielak}, {Rosner}, \&
  {Ulmschneider}}]{musielak+al1989}
{Musielak}, Z.~E., {Rosner}, R., \& {Ulmschneider}, P. 1989, \apj, 337, 470

\bibitem[{{Musielak} {et~al.}(2000){Musielak}, {Rosner}, \&
  {Ulmschneider}}]{musielak+al2000}
---. 2000, \apj, 541, 410

\bibitem[{{Musielak} \& {Ulmschneider}(2001)}]{musielak+ulmschneider2001}
{Musielak}, Z.~E., \& {Ulmschneider}, P. 2001, \aap, 370, 541

\bibitem[{{Parker}(1974)}]{parker1974}
{Parker}, E.~N. 1974, \apj, 189, 563

\bibitem[{{Rae} \& {Roberts}(1982)}]{rae+roberts1982}
{Rae}, I.~C., \& {Roberts}, B. 1982, \apj, 256, 761

\bibitem[{{Rezaei} {et~al.}(2007){Rezaei}, {Steiner}, {Wedemeyer-B{\"o}hm},
  {Schlichenmaier}, {Schmidt}, \& {Lites}}]{rezaei+al2007}
{Rezaei}, R., {Steiner}, O., {Wedemeyer-B{\"o}hm}, S., {Schlichenmaier}, R.,
  {Schmidt}, W., \& {Lites}, B.~W. 2007, \aap, 476, L33

\bibitem[{{Rimmele}(2004)}]{rimmele2004}
{Rimmele}, T.~R. 2004, \apj, 604, 906

\bibitem[{{Rosenthal} {et~al.}(2002){Rosenthal}, {Bogdan}, {Carlsson}, {Dorch},
  {Hansteen}, {McIntosh}, {McMurry}, {Nordlund}, \& {Stein}}]{rosenthal+al2002}
{Rosenthal}, C.~S., {et~al.} 2002, \apj, 564, 508

\bibitem[{{Schaffenberger} {et~al.}(2005){Schaffenberger},
  {Wedemeyer-B{\"o}hm}, {Steiner}, \& {Freytag}}]{schaffenberger+al2005}
{Schaffenberger}, W., {Wedemeyer-B{\"o}hm}, S., {Steiner}, O., \& {Freytag}, B.
  2005, in ESA Special Publication, Vol. 596, Chromospheric and Coronal
  Magnetic Fields, ed. {D.~E.~Innes, A.~Lagg, \& S.~A.~Solanki} (Noordwijk:
  ESA)

\bibitem[{{Schaffenberger} {et~al.}(2006){Schaffenberger},
  {Wedemeyer-B{\"o}hm}, {Steiner}, \& {Freytag}}]{schaffenberger+al2006}
{Schaffenberger}, W., {Wedemeyer-B{\"o}hm}, S., {Steiner}, O., \& {Freytag}, B.
  2006, in ASP Conf.~Ser., Vol. 354, Solar MHD Theory and Observations: A High
  Spatial Resolution Perspective, ed. {J.~Leibacher, R.~F.~Stein, \&
  H.~Uitenbroek} (San Francisco, CA: ASP), 345

\bibitem[{{Schrijver} {et~al.}(1989){Schrijver}, {Cote}, {Zwaan}, \&
  {Saar}}]{schrijver+al1989}
{Schrijver}, C.~J., {Cote}, J., {Zwaan}, C., \& {Saar}, S.~H. 1989, \apj, 337,
  964

\bibitem[{{Simon} \& {Leighton}(1964)}]{simon+leighton1964}
{Simon}, G.~W., \& {Leighton}, R.~B. 1964, \apj, 140, 1120

\bibitem[{{Skumanich} {et~al.}(1975){Skumanich}, {Smythe}, \&
  {Frazier}}]{skumanich+al1975}
{Skumanich}, A., {Smythe}, C., \& {Frazier}, E.~N. 1975, \apj, 200, 747

\bibitem[{{Steiner} {et~al.}(1998){Steiner}, {Grossmann-Doerth}, {Kn\"olker},
  \& {Sch\"ussler}}]{steiner+al1998}
{Steiner}, O., {Grossmann-Doerth}, U., {Kn\"olker}, M., \& {Sch\"ussler}, M.
  1998, \apj, 495, 468

\bibitem[{{Vigeesh} {et~al.}(2009){Vigeesh}, {Hasan}, \&
  {Steiner}}]{vigeesh+al2009}
{Vigeesh}, G., {Hasan}, S.~S., \& {Steiner}, O. 2009, \aap, 508, 951

\bibitem[{{Ziegler} \&
  {Ulmschneider}(1997{\natexlab{a}})}]{ziegler+ulmschneider1997a}
{Ziegler}, U., \& {Ulmschneider}, P. 1997{\natexlab{a}}, \aap, 324, 417

\bibitem[{{Ziegler} \&
  {Ulmschneider}(1997{\natexlab{b}})}]{ziegler+ulmschneider1997b}
---. 1997{\natexlab{b}}, \aap, 327, 854

\end{thebibliography}

\end{document}